\documentstyle[epsfig]{aipproc}

\begin{document}
\begin{flushright}
 June  12  1997\\
TAUP 2431/97\\
\end{flushright}

\title{In combat with nonperturbative QCD}
\author{ Eugene   Levin }
\address{  School of Physics and Astronomy\\
 Raymond and Beverly Sackler Faculty of Exact Science\\
 Tel Aviv University, Tel Aviv, 69978, ISRAEL\\
and\\
 Theory Department, Petersburg Nuclear Physics
Institute\\
 188350, Gatchina, St. Petersburg, RUSSIA}

\maketitle

\begin{abstract}
This talk is the second part of the summary of the theory section at
DIS'97. We 
demonstrated  that  theory of the ``hard" processes,  based on QCD, has
achieved  a better understanding of available experimental data during the
past year. This was due to  the remarkable success achieved in
calculating of high order corrections in perturbative QCD ( see the
summary talk of S. Forte) as well as 
 progress in nonpertutbative QCD approaches.

\end{abstract}
\section*{Introduction}
In this talk we  briefly  outline the main ideas,
approaches and results of the calculations of  the ``hard'' processes that
were discussed in the theory working group at DIS97. The main message,
which we wish to convey to a reader, is that  the development of theory
during the past year  has been exciting and solid, mostly because of
remarkable breakthrough in the calculation of the  high orders in
perturbative QCD (pQCD) ( see the summary of S. Forte)  and because of the
new ideas and methods for the
nonperturbative contributions in QCD (npQCD). To have a better
understanding of what was  achieved, we  start with basics of our
knowledge on QCD. 

Everyone knows, that:

1. ``hard" processes occur at small distances and they are the most
suitable  ones to
apply the pQCD approach especially
for such inclusive observables as the deep inelastic structure functions
or/and the cross sections of ``hard" production;

2. for ``hard" processes we can use a very general formalism of the
renormalization group approach which gives a deeper insight in the main 
properties of QCD than obtained from any sophisticated calculations in
pQCD;

3. the renormalization group approach leads to the DGLAP \cite{DGLAP} 
evolution equations, which play a role of the Coulomb law in QCD;

4. for ``hard" processes we have a regular procedure  which  takes into
account
a power - like corrections with respect to the ``hard" scale, so called
Wilson Operator Product Expansion ( WOPE))\cite{OPE};
 
5. for ``hard" processes the factorization theorem was proven \cite{FACT},
which allows us to separate the ``hard" and ``soft" contributions.

All these properties make the ``hard" processes a good training ground for
all theoretical methods. Unfortunately, the main theoretical  method for
``hard"
processes is still pQCD. However, if one wants to characterize all
contributions  to our section in one sentence, it is {\it we have to  
start
to discuss   nonperturbative contributions to DIS}. These
nonperturbative approaches have been widely presented in our section as
well as in proceedings, and we tried to include not only the results of
the past year 
 but also the minireviews of the situation. We are happy to report
that methods of npQCD  such as renormalons, instantons, semiclassical
field
approach, high parton density QCD and lattice calculations have been
covered in  detail, giving the uptodate picture of the situation in
npQCD at small distances. 

The goal of this talk is to review the status of the 
understanding of nonperturbative QCD at small distances,, that has been
reached in our working
group.
\section*{ Perturbative QCD}
The  goal of this section is to recall the main ideas and terminology  of
pQCD to help  the 
reader in understanding our strategy in dealing with the nonperturbative
 contributions to the ``hard" processes. One can find  an uptodate
review of the progress in pQCD in the Forte's talk \cite{FORTE}.
\subsubsection*{Two devils, that we are struggling  with}

In DIS, where the
typical  transverse distances ($ r_{\perp}$) are small,  we have a
natural small parameter, $\alpha_S(r^2_{\perp})\,\ll\,1$. Therefore, at
first sight, we have to calculate a couple of the Feyman diagrams to
obtain the answer. However, the situation  is much more
complicated and, doing the pQCD calculation, we are always struggling with
two major problems: 

1. the real parameter of pQCD is not
$\alpha_S(r^2_{\perp})$ , but  $\alpha_S(r^2_{\perp})\,L$, where $L$ is a
large log. For example, a gluon structure function $xG(x,Q^2)$ can be
written as a perturbative series in the form:
\begin{equation}
x G(x,Q^2)\,\,=\,\,\sum^{\infty}_{n = 1}\,C_n  \alpha^n_S\,(\, L^n\, +\,
a_{n-1} L^{n - 1}\,+\,...\,a_0\,)\,\,,
\end{equation}
where   $L$ could be:
\begin{equation}
1.\,\,\,L\,=\,\ln(Q^2/Q^2_0)\,\,\,\,\,at\,\,Q^2\,\gg\,Q^2_0\,\,\,and\,\,
\,x\approx\,1\,\,;
\end{equation}
$$
2.\,\,\,L\,=\,\ln(1/x)\,\,\,\,\,at\,\,Q^2\,\approx\,Q^2_0\,\,\,and\,\,
\,x\,\rightarrow\,0\,\,;
$$
$$
3.\,\,\,L\,=\,\ln(Q^2/Q^2_0)\,\ln(1/x)\,\,\,\,\,at\,\,Q^2\,\gg\,Q^2_0\,\,\,
and\,\,
\,x\rightarrow\,0\,\,;
$$
$$
4.\,\,\,L\,=\,\ln(1 - x)\,\,\,\,\,at\,\,Q^2\,\approx\,Q^2_0\,\,\,and\,\,
\,x\rightarrow\,1\,\,.
$$
Therefore, to obtain  first approximation to a physical observable we
have to sum all contributions of the order $(\alpha_s\,L)^n$, not only a
couple of diagrams. This is a rather difficult technical problem, which we
 usually call resummation.

2.The second and more serious problem is  the fact
that $ C_n \,\rightarrow\,n!$ at large values of $n$. This  means that
this
series is the asymptotic one. We have no general approach for a summation
of such a series,  occasionally, we can guess an
analytic
function which has the same perturbative series. Sometimes, we develop
a general approach for summation ( mostly when $C_n\,\rightarrow \,(-1)^n
\,n!$ ), but mostly it is an open problem. We know, at the moment, at
least
three sources  for $n!$ behaviour: infrared and ultraviolet renormalons
and
instantons. All three give us a window to npQCD and we will discuss them
later. Our practical strategy of dealing with asymptotic series is  as
follows:

1. Instead of the full series of Eq.(1), we  first sum a simpler series,
so
called, leading log approximation (LLA):
\begin{equation}
xG^{LLA}(x,Q^2)\,=\,\sum^{\infty}_{n=1}\,C_n\,(\alpha_s\,L)^n\,\,,
\end{equation}
for which we have evolution equations. Solving them we have an analytic
function. Actually, there are  three evolution equations for the deep
inelastic structure functions  on the market: the
celebrated  Dokshitser - Gribov - Lipatov -Altarelli - Parisi (DGLAP)
 evolution equation \cite{DGLAP} for region 1 in Eq.(2), the Balitsky
- Fadin - Kuraev - Lipatov  equation or so called the BFKL Pomeron\cite{BFKL}
 for
region 2 in Eq.(2) and the Ciafaloni-Catani - Fiorani - Marchesini (CCFM)
evolution equation \cite{CCFM} which provides a correct matching between
region 1 and
region 2 .  The current situation with the GLAP evolution equation is
reviewed in Ref. \cite{FORTE}, while the other two we will discuss here.
We want to emphasize that we need to develop the LLA approach for each
``hard"process with the careful analysis of the value of scale $L$ in
different kinematic regions. This is a very difficult technical problem,
in which  remarkable  progress has been made ( see Forte's talk
\cite{FORTE}).

2. We calculate the ratio
\begin{equation}
R(x,Q^2)\,\,=\,\,\frac{x
G(x,Q^2)}{xG^{LLA}}\,\,=\,\,\sum^{\infty}_{n=1}\,r_n\,\,=\,\,\sum^{\infty}_{n=1}
C_n( L^{n - 1}\,+\,...+\,a_0)\,\,.
\end{equation}
The  situation today in computation of the terms $r_n$ in Eq.(4)  is
reviewed
in Ref. \cite{FORTE}, in   most processes we know $r_1$ and $r_2$,
sometimes even $r_4$ has been calculated.
 
3. Our hope is that $\frac{r_n}{r_{n - 1}}\,\ll\,1$ for sufficiently large
$n\,=\,N$.

4. The result of the calculation should be given in the form:
\begin{equation}
R(x,Q^2)\,\,=\,\,\sum^{n = N - 1}_{n=1}\,r^n\,\,\pm\,\,r_N\,\,.
\end{equation}
Therefore, inside  pQCD we have an intristic accuracy which  makes
 all calculations difficult . Practically, Eq.(5) means that,
strictly
speaking, we have to use  calculations in the next order in
$\alpha_s$,
only as an error for the result of calculation in the previous order.

5. The accuracy of the pQCD approach depends crucially  (i)on our skill in
defining  the value of scale $L$ for the process of interest, (ii) on 
the value of $N$ for particular process, (iii)  on our success in
solving
the LLA equations and  (iv) on  our understanding of $n!$ behaviour of
$C_n$ in our perturbative series.

\subsubsection*{The factorization theorem}

Any calculation of ``hard" processes is  based on the factorization
theorem \cite{FACT}, which allows us to separate the nonperturbative 
contribution from large distances (parton densities,  $F^i_A(\mu^2)$) from
the
perturbative one (``hard" cross section, $\sigma^{hard}$). For example,
the cross section
of the high $p_t$ jet production in hadron - hadron collisions
 can be written schematically in the form:
\begin{eqnarray}
\sigma( A + B \rightarrow jets(p_t) + X)\,\,=\nonumber\\
\,\,F^i_A(\mu^2)\, \bigodot
\,F^i_A(\mu^2)\,\bigodot\,\sigma^{hard}(partons \,\,with
\,\,p_t\geq k_t\geq \mu)\,\,.
\end{eqnarray}
As we  have  discussed, we need to calculate $\sigma^{hard}$  not
only in
the leading order of pQCD but also in high orders so as  to specify the
accuracy
of our calculation. Practically, we need to calculate the ``hard" cross
section  at least in the next to leading order,  to reduce the scale
dependence, which appears in the leading order calculation, as a clear
indication of the low accuracy of our calculation. Of course, we have to
adjust  the accuracy in the calculation of $F(\mu^2)$ and $\sigma^{hard}$.
Solid progress has been achieved in this program \cite{FORTE}.

\section*{On the border between perturbative  and nonperturbative QCD}

\subsubsection{The BFKL and CCFM  equations}
The BFKL equation was derived in LLA of  pQCD summing $\alpha_S
\ln(1/x))^n$
terms in the kinematic region   ( see region 2 in Eq.(2))  without any
``hard" scale \cite{BFKL}.
 The simple physics behind the infrared
stable answer,  related to new
degrees of
freedom at high energy ( colour dipoles \cite{MU94}), as well as the 
beauty of
hidden symmetries, make this equation a hot subject of investigation
for
theorists, as a possible model of matching  of the ``hard" and ``soft"
processes in QCD. The fact that the BFKL dynamics has not been seen in
HERA data, that have penetrated the BFKL kinematic region, brought an
additional challenge to  experts. For a long time the BFKL approach 
sufferred from the lack of  corrections in the next to leading
order,
without which it was impossible to estimate the accuracy of
the calculation.

The {\bf hot and good news} of this conference is that these corrections
were
calculated and presented for the first time in our section \cite{FADIN}.
 The bad news is the fact that we have not understood these corrections,
and we are still not able to tell  how they will change the
energy dependence of the cross section and the value of the anomalous
dimension ( $Q^2$ dependence) in DIS.  The expectation is that they lead
to a smoother energy behaviour and to an  increase of the anomalous
dimension. However,unfortunately, we have to postpone this discussion till
DIS98. 

Fortunately, we do know what to look for in the calculations of
Ref.\cite{FADIN}. The beauty of the BFKL equation is based on two simple
properties of the leading log (1/x) approximation  of pQCD: the separation
of the longitudinal and transverse degrees of freedom,,,,, and conformal
symmetry related to dimensionless coupling in QCD. Both of them are broken
in the next order. The running coupling constant brings the scale to the
physical observables  and
separation of the longitudinal and transverse degrees of freedom does not
hold. However, one can see  that   the gluon BFKL kernel of the next to
leading order
(see Ref.\cite{FADIN}) as well as the quark one \cite{CIAFALONI} confirms
the suggestion \cite{BRA} that we need to substitute $\alpha_S\,
\rightarrow \frac{\alpha_S(q^2_1)\alpha_S( (q_1 -
q_2)^2)}{\alpha_S(q^2_2)}$ in the leading order BFKL kernel to take into
account the running coupling constant of QCD. $q_1$ is the transverse
momentum of radiating gluon, $q_1 - q_2$ is the transverse momentum of
produced gluon, while $q_2$ is the transverse momentum of the gluon after
emission. The second way of introducing the scale into the BFKL
equation is
closely related to kinematic constrain taken  into account in the CCFM
equation ( see Ref. \cite{SALAM} ). The CCFM equation is an elegant  
way of
including in the equation for the parton density  the angle ordering in
the parton cascade which holds in any order of the perturbation theory.

Therefore, we  hope that if we  write the CCFM equation instead of
the BFKL one \footnote{It should be stressed that the CCFM equation leads
to the BFKL one in the LLA.}, and if we include the running QCD coupling
as
has been described above, the rest of the kernel will preserve 
conformal invariance. On the other hand, we have some arguments that the
conformal invariant part of the next to leading BFKL kernel has a very
definite form \cite{WHITE}.

We  notice that  real progress in the CCFM equation has
been achieved not only in the  matching of two kinematic regions for the
deep
inelastic structure function but in creating a theoretical approach to
multigluon production \cite{SALAM}. It gives one  hope of building  a
theoretical description of the inclusive processes in the BFKL - like
dynamics.

Most  experts  feel that the BFKL equation is more fundamental than the
LLA
of pQCD in which it was originally proven. This is the reason why we
return
 to the proof of the BFKL equation using different approaches. It is
well known that this strategy has produced results, namely, the
new degrees of freedom ( colour dipoles \cite{MU94} ) at high energy
interaction. In our working group we discussed four new derivations:
in string - based method of pQCD \cite{DELDUCA}, in new renormalization
group approach to longitudinal degrees of freedom (Ref. \cite{BALL} and
below), in the OPE - like formalism for high energy scattering
\cite{BALITSKY}
and in semiclassical field approach (Ref. \cite{KOVNER} and below). In
Ref.\cite{BALITSKY} a  new formalism was suggested which gives the OPE
for
high energy scattering (HEOPE). In the HEOPE,   power-like
corrections in energy squared  $s$ has been separated  and assigned to
nonlocal
operators (Wilsons lines). Each  nonlocal operator has only $\ln s$
dependence on energy and the first one ( one Wilson line) has the energy
dependence given by the BFKL equation. This approach passed the first non
trivial test: in Ref. \cite{BALITSKY} the so called triple Pomeron vertex
was calculated and the answer coincides with the expression obtained in
pQCD ( see Ref. \cite{BAWU}). Much more work  needs to be done  to
understand how
general this approach is, but if it is a correct one, we have a chance
to 
understand better   the high energy asymptotic beyond of pQCD.
 
\subsubsection{  A violation of the factorization theorem at high energy}

We have two examples that the factorization theorem, which is the basis of
all our  calculations of the ``hard" processes, has only limited accuracy
at high energy ( low $x$).

The first one was given in Ref.\cite{MU97}, where was shown that, 
due to the BFKL dynamics, the WOPE
 can only  be  used safely for DIS at low  $x$ ( say $ x_0 < x \ll 1$)
 in
the limited region of $Q^2$, namely  for
 $Q^2 \,\geq\, Q^2(x_0)$ where
\begin{equation} 
\ln (Q^2(x_0)/\Lambda^2)\,\geq \,[ \frac{7 N_c \zeta(3)}{\pi 
b}\,\ln(x_0/x)]^{\frac{1}{3}}\,\,,
\end{equation}
  $\zeta$ is Riemann zeta function and
all other notations are clear from $\alpha_s(q^2)\,=\,\frac{4 \pi}{b
\ln(q^2/\Lambda^2)}$. However, what is more important that this breakdown
of the WOPE is not due to higher twist terms but rather to an inability
to properly separate ``hard" and ``soft" scales at $x \,\rightarrow\,0$.
This  means that the factorization theorem does not work at high energies.
 This   statement has a very simple physical meaning. Indeed, in the
BFKL kinematic region the mean transverse momentum of partons 
increases with energy. Therefore, we have a two scale problem even for
the
deep inelastic structure function. Only at a high value of $Q^2$( given by
Eq.(7)) when $Q^2$ is bigger than the mean parton transverse momentum,
 $Q^2$ become the obvious  scale of hardness in DIS and we can use the
WOPE.

The second example was discussed for the first time in our working group
\cite{LEVIN}. It was argued that some interference diagrams have been
missed in the proof of the factorization theorem \cite{FACT} these  will
give a sizable contribution for high energy `` hard" inclusive production,
especially for nucleus collisions.

Both of these examples show that we have to reconsider the formal proof of
the factorization theorem to determine a correct region of its
applicability. 

\subsubsection{ Factorization for ``hard" exclusive processes}

As was  recently shown  \cite{RY} \cite{FIVE} the diffractive
electroproduction of vector mesons can be  treated in pQCD ( see Ref.
\cite{CFS} for a general proof), since  all nonperturbative
contributions can be factorised out in the  wave function
 of the produced vector meson. In Ref. \cite{CFS} it was shown, that the
amplitude of diffractive production ( $M$) of the vector mesons as well
as
other hadrons can be written in the factorizable form:
\begin{eqnarray}
M (\gamma^* + A\,\rightarrow\,V + A)\,\,=\\
F^i_A ( x_B,x_M,\mu^2)\,
\bigodot\,H_{ij}(Q^2,..)\,\bigodot\,\varphi^V_j(\mu^2)\,\,
+\,\,O(\frac{\mu^2}{Q^2})\,\,,\nonumber\\
\end{eqnarray}
 where $H_{ij}$ is the ``hard" scattering function, $\varphi$ is the
light-cone wave function of  the meson \cite{BL} and $F^i_A (
x_B,x_M,\mu^2)$
is the off-diagonal parton density.The new kinematic variable      $x_M$
is
equal to  $ 
M^2/W^2$.

The
off-diagonal parton densities  were actually  introduced  long ago
in
Refs.\cite{BALO} \cite{GLR} where the diffractive production of Z in
DIS was
considered. The difference, between  the off-diagonal parton densities and
the
conventional ones,  is   due to the fact that the longitudinal momentum
transfer in the diffractive production in DIS is not equal to zero, as it
is 
 for the conventional deep inelastic structure functions. For the off -
diagonal parton densities we have a new evolution equation which was
suggested first in Ref. \cite{GLR} ( see also \cite{GER}). In our section
two papers were
presented \cite{RAD} \cite{FRE} where these new evolution equations were 
 re-derived, checked, corrected and generalized. We think, that this  is
a 
closed chapter in the  leading order of pQCD ( see Ref.\cite{RAD} for the
most detail and rigorous proof).

\subsubsection{High twist contribution}
 Due to the WOPE, a deep inelastic structure function can be written in
the form:
\begin{equation}
F_2(x,Q^2)\,=\,F^{LT}_2(x,Q^2 )\,\,+\,\,\frac{m^2}{Q^2}\,F^{HT}_2\,\,.
\end{equation}
We know  almost everything about $F^{LT}_2$ where the DGLAP
evolution equations apply. We know a lot about the next order twist
structure function
 $F^{HT}_2$: the physical meaning \cite{ELLISHT}, the evolution equations
\cite{BULI} and even the solution of the evolution equations in the region
of low $x$ \cite{HTLOWX}.  We know enough to look with smile,  at numerous
attempts of experimentalist to fit the data assuming that $F^{HT}_2 $
has the same $Q^2$ dependence as $F^{LT}_2$. However, nobody has yet given
a numerical
estimate of $Q^2$ and $x$ dependence of the high twist contribution based 
on what we know.
This has been done \cite{BARTELSHT} and the result is  to some extent 
surprising. The extra power of $Q^2$ does not give the  feeling
that this
contribution should be negligibly small at least at $Q^2\,\approx\,10
GeV^2$. Indeed, if one  wants to try a simple parameterization for
$F^{HT}_2$
it is better to take $F^{HT}_2\,\propto\, ( F^{LT}_2)^2 )$ \cite{HTLOWX}
\cite{BARTELSHT} and the $Q^2$ dependence due to anomalous dimension of
$F^{LT}_2$ compensates to large extent for  the $1/Q^2$ suppression. The
importance of this result is obvious, since in all solutions of the DGLAP
equations that there are on the market, it has been assumed that  the
higher
twist contribution is small and can be neglected at $Q^2 = Q^2_0$ where
$Q^2_0$ is the initial virtually of the photon from which we start the
DGLAP evolution. Notice that in practice the value of  $Q^2_0$ is 
rather small ( about 4 $GeV^2$ ).
  
\subsubsection{Back to shadowing corrections}
The HERA data on deep inelastic structure function lead
to puzzling result. On one hand they can be successfully described in
the framework of the DGLAP evolution equations without any new ingredients
like the BFKL equation and/or the shadowing corrections (SC). On the other
hand  the parameter ($\kappa$) which gives the estimate for the strength
of
the SC turns out to be large  ($\kappa\,\geq\,1$) in the HERA kinematic
region. This parameter $\kappa$ was estimated in Refs.\cite{GLR}
\cite{MUQI} and it is equal to
\begin{equation}
\kappa\,\,=\,\,xG(x,Q^2)\,\frac{\sigma(GG)}{\pi R^2}\,\,=\,\,x
G(x,Q^2)\,\frac{3 \pi \alpha_s}{Q^2\,R^2}\,\,.
\end{equation}
To understand what is going on, we have to develop a theoretical approach
in which we can treat the region of $\kappa \,\approx\,1$ in DIS. It
should
be stressed that the previous attempts to develop a theory for the SC
\cite{GLR} only had
a guaranteed theoretical accuracy  for small $\kappa
\,\approx\,\alpha_s\,\ll\,1$. Two of such approaches were discussed in the
working group: in the first one\footnote{E.Levin gave a talk on this
subject but he did not write a contribution to the proceedings.} 
\cite{AGL}     pQCD was used at the edge of its validity
($\alpha_s \kappa\,\leq\,1$), while in the second
( see Refs.\cite{WEIGERT}\cite{KOVNER}and below ) the new approach was
developed  in the kinematic region of  high parton density QCD.

In Ref. \cite{AGL} a new evolution equation was derived which describes
 that each parton   in the parton cascade interacts with the target in
Glauber - Mueller approach \cite{MU90}.  The results are the following:
(i)
$\kappa$ is the correct parameter that determines  the strength of the SC;
(ii) the SC   to the gluon structure function  are big   even in the  HERA
kinematic region, but nevertheless  the value of the shadowed gluon
structure function   is still within the experimental
errors or, another way of putting it, the difference between the shadowed
and
nonshadowed gluon structure functions does not exceed the difference
between the gluon structure functions in the different parameterizations
such as the MRS,GRV and CTEQ ones; (iii) the SC to $F_2(x,Q^2)$ 
 in HERA kinematic region are so small  that can be neglected;  (iv) the
SC enter
the game before the BFKL equation and, therefore, the BFKL Pomeron cannot
be seen in the deep inelastic structure function  since it is hidden
under substantial SC; and
  (v) in the
region of low $x$ the asymptotic behaviour of $xG(x,Q^2)$ is $xG(x,Q^2)\,
\rightarrow\, \frac{2\, R^2\, Q^2}{ \pi^2} \,\ln(1/x)\,\ln \ln (1/x)$.
This
means that the gluon  density does not saturate \cite{GLR}
unlike in the GLR equation.
\subsubsection{New ideas}
 A new idea of  how to calculate the high energy amplitude in QCD was
presented in Ref. \cite{BALL}: i.e.  to use the renormalization group
approach
for the longitudinal degrees of freedom. The arguments are based on the
$k_t$ -
factorization \cite{KTFACT} and on similarity between $\ln Q^2$ and $\ln
s$. The answer is, roughly speaking, the BFKL amplitude, but with running
QCD coupling which depends on energy. Certainly, this answer does not
contradict  unitarity and, perhaps, even the experimental data, but, of
course, it is in strong contradiction with the BFKL approach, that has
been discussed in subsection 6.   During the discussion in our working 
group we
did not reach  agreement and more time is needed to
clarify the situation. However, the fact that none  of the experts 
has seen    a diagram  which survives at high energy and in which the
running
coupling constant depends on $s$, makes the whole approach rather
suspicious.  On the other hand, this approach is sure to stimulate   a
more
detailed study of this problem.
\section*{ Real nonperturbative stuff}
\subsubsection{ Semiclassical gluon field approach}
The real nonperturbative contributions are closely related to  the ``soft"
gluonic field with  small transverse momenta ( $q_t\,\leq\,\Lambda$) and
small frequency ($\omega_q\,=\,q_{-}\,=\,\frac{q^2_t}{q_{+}}\,\leq
\,\Lambda$) where $\Lambda$ is QCD scale ($ \alpha_s (k^2)\,=\,\frac{4
\pi}{b\ln(k^2/\Lambda^2)}$). This is an old idea \cite{BJ}\cite{NACHT},
which led to certain understanding of nonperturbative behaviour of the
deep inelastic structure functions. What is new  is that we  can say a lot
about ``hard" processes (DIS), using     only a general form of the QCD
Lagrangian \cite{BUCH}.  The physical picture of this approach is
extremely transparent: the colour dipole of a small size ($r_t
\,\approx\,\frac{1}{Q}$ for DIS) penetrates the cloud of  the  ``soft"
gluon
field which can be treated semiclassically. The amazing result is that we
can
obtain  very definite predictions without a detailed knowledge of the
structure of  this cloud. For example, this picture leads  to the
following prediction for  diffractive charm production in DIS:
$$ \sigma^{DDcharm}_L
\,\approx\,\frac{1}{Q^2}\,\,\,\,\,\sigma^{DDcharm}_T\,\approx\,
\frac{\ln(Q^2/m^2_c)}{Q^2}\,\,,$$ 
which have  to be confronted with the
pQCD results\cite{LMRT}:
$$ \sigma^{DDcharm}_L
\,\approx\,\frac{\Lambda^2}{Q^2}\,\ln(Q^2/m^c)\,\,\,\,\,
\sigma^{DDcharm}_T\,\approx\,
\frac{\Lambda^2}{Q^2 \,m^2_c}\,\,.$$ 
In Ref.\cite{BUCH} one can find more predictions for future
experiments which will help to differentiate the nonperturbative and
perturbative contributions to DIS. 
   
It turns out that this idea of semiclassical gluon field is very useful in
obtaining the effective Lagrangian for high parton density QCD
( see Refs. \cite{MCLV},\cite{WEIGERT} and \cite{KOVNER} ). The physical
problem has been  pointed out a long ago ( see Ref. \cite{GLR}and
Ref.\cite{LALE} for updated review): at high energy (low
$x$ ) and /or for DIS with heavy nucleus we are dealing with the system of
partons so dense that conventional methods of pQCD does not work. However,
the typical distances are still small for DIS    and this fact results in
weak correlations between partons due to small the  coupling constant of
QCD.

The revolutionary idea, suggested in Ref. \cite{MCLV} and developed in
Refs. \cite{WEIGERT} and \cite{KOVNER}, is: in the Bjorken frame for DIS 
we can replace the complex QCD
interaction between parton in such a system by the interaction of a parton
($i$)
with energy fraction $x_i$ with the classical field created by all partons
with energy fraction $x$ bigger than $x_i$. Indeed, in leading log(1/x)
approximation of QCD  all parton with $x\,>\,x_i$ live for a  much longer
time
than parton $i$, therefore, they create a gluon field which only depends
 on
their density. Using this idea and Wilson renormalization group approach,
in  Refs.\cite{WEIGERT} and  \cite{KOVNER} ( see also Ref.\cite{KOVCH}
for elucidating  remarks), the effective Lagrangian was
obtained. 

It has been demonstrated that this effective Lagrangian
correctly reproduces the DGLAP evolution equations \cite{WEIGERT} and even
the BFKL Pomeron \cite{KOVNER} in the limit of a sufficiently weak gluon
field. However, the main problem in  matching the  two approaches: one
which
we  discussed in subsection 5 and this one, is still an open problem.
The equation which was originally obtained in Ref.\cite{WEIGERT}, does not
match, but in our working group we were instructed by A.Kovner and
H.Weigert that the equation is  not complete. We hope that this problem
will be solved soon and we will have a reliable theoretical approach for
high
parton density QCD which will allow us to search for new collective
phenomena in QCD at HERA, Tevatron, RHIC and at LHC.

\subsubsection{Renormalons and Instantons}
As mentioned previously, renormalons and instantons are the known sources
of
the $n!$ behaviour of coefficients of the perturbative series which give
 estimates for the value and $Q^2$ dependence of $r_N$ term in Eq.(5).

In
our working group we covered the renormalon contributions in much detail
 ( see Forte's talk
\cite{FORTE} for a comprehensive review of the present situation). For the
sake of completeness I list here the main
achievements that have been made: (i) a successful phenomenology based on
renormalons as the way to take into account  nonperturbative
corrections
\cite{BENEKE} \cite{AKHOURY} \cite{MARCHESINI}; (ii) confirmation of the
WOPE in  processes with one scale of hardness and the possibility to
calculate power - like corrections with respect to hard scale ($Q^2$) to
processes without the WOPE \cite{BENEKE} \cite{MARCHESINI}; (iii)a  
powerful
dispersion relation method which allows us to treat the nonperturbative
corrections to a variety of processes; (iv) the general nonlocal OPE for
processes without the WOPE \cite{STERMAN} and (v) the possibility to
discuss corrections to the WOPE    due to high order terms in the
perturbative series \cite{MARCHESINI}.

Renormalons give us the possibility to discuss  power-like
nonperturbative corrections to
the BFKL equation (see Refs. \cite{BRA} \cite{SOT} ). It turns out that
the nonperturbative corrections from infrared instantons have different
dependences on $Q^2$ for the gluon structure function ($\frac{1}{Q}$
\cite{BRA} \cite{SOT}), and
for $F_2(x,Q^2)$ ( $\frac{1}{Q^2}$\cite{SOT}).

Instantons were reviewed in Ref.\cite{SCHREMPT}. The main result is that
the instanton contributions can be theoretically  calculated in
inclusive production in  DIS, and an experimental observation of a
typical instanton event in the DIS, which has a clear signature, will open
a new insight regarding  the nonperturbative contribution. There is no QCD
without
instantons, therefore, an experimental observation of an instanton event
is a great challenge to experimentalists.

\subsubsection{Lattice calculation}
For the first time  the lattice calculation in DIS  were discussed
seriously.
The experimental errors in the lattice experiment are  still large  but,
nevertheless, it gives a convincing result that the initial quark
distributions at $Q^2 = 2.5\,GeV^2$ differs from experimental one ( see
minireview \cite{LATTICE}). Such a difference was expected since in the 
lattice experiment the leading twist contribution has been calculated
while  experimental data give the contribution of all twists at
definite value of virtuality $Q^2$. It is interesting to note that the
 leading twist quark distributions derived on the lattice turn out to be
closer to 
one that was expected in the constituent quark model, where the mean
momentum
of
the quark   about $\frac{1}{3}$. 

We would like to emphasize that the lattice calculations give an unique
opportunity to develop a self consistent theoretical approach to  DIS. 
The following strategy is advocated . We first use the lattice parton
leading
twist densities as the initial conditions for the GLAP evolution
equations and  solve them. The difference between experimental initial
parton densities and the lattice one should be treated  as the
high twist
contribution. For them we should use the high twist evolution equations
and  the theoretical  status of the high twist contribution  was
discussed.
The above procedure will provide a theoretical  approach to  DIS
and
after it has been achieved  we will be able to discuss DIS on the
solid
theory basis.  However,  the experts in lattice 
experiments need  to calculate the initial gluon density which  has not
 yet been done.
  
\section*{Conclusions}
Going  back over  the titles of our subsections one can see the main
subjects  of
our discussions. The first trivial conclusion is that the theory of
``hard" processes  is in a
very good shape in the  year 1997. Of course, we are still struggling  
with
nonperturbative QCD but, obviously, we are in offensive phase, attacking a
number of very difficult problems and making    great progress in our
understanding. An illustration of this could be our recommendation for
phenomenology of ``hard" processes. They are:

1. the calculation of the next order corrections to the BFKL equation
which will
allow us to calculate the accuracy of the new phenomenological approach
based on the BFKL anomalous dimension \cite{THORNE} which provides a
reasonable matching  of ``hard" and ``soft" processes in QCD;

2. the theoretical discussion of the high twist contribution (see
subsection 4) suggests
  the following formula to fit the deep inelastic
structure
function:
$$
F_2(x,Q^2)\,\,=\,\,F^{DGLAP}_2(x,Q^2)\,\,+\,\,\frac{m^2}{Q^2}c(x)\,
[\,F^{DGLAP}_2(x,Q^2)\,]^2\,\,,
$$
where $c(x)$ is a phenomenological function of $x$;

3. using lattice calculations we can develop a self consistent theoretical
approach to predict the deep inelastic structure function ( see subsection
9 );

4. the semiclassical gluon field approach ( see subsection  7) gives a
definite prediction for diffractive charm production which should be
checked experimentally;

5. new studies on shadowing corrections (see subsections 5 and 7) show
the urgent need to determine experimentally, or extract
phenomenologically, the
gluon parton density;

6. the factorization theorem has only a limited region of applicability
and
should be used with  caution.

The last remark that we wish to make is,, that  
the
new physics beyond the Standard Model    was not discussed here,
 as  this item  was
discussed in detail in the structure function working group. Our
theoretical speakers ( in our common session with the structure function
WG ) demonstrated that  new particles like leptoquark
are needed  in theory and, perhaps to the  surprise of the non experts,
the
theory is  now in  such  good shape that it can predict a lot, in
particular, the manifestation of the same mechanism which could lead to
new
HERA data on high $Q^2$ production in a variety of other processes.
Welcome to DIS'98 for more profound discussions on this issue, if at all.

{\bf Acknowledgements:} I am grateful to Stefano Forte for pleasant and
fruitful cooperation in conducting theory DIS'97. I am thankful to all
members of our WG for their contributions and discussions which make our
work a success. Special thanks  to Asher Gotsman for his comments on the
manuscript.


\begin{thebibliography}{99}
\bibitem{DGLAP}
V.N. Gribov and L.N. Lipatov:{\it Sov. J. Nucl. Phys.} {\bf 15} (1972)
438; L.N. Lipatov: {\it Yad. Fiz.} {\bf 20} (1974) 181; G. Altarelli and
G. Parisi:{\it Nucl. Phys.} {\bf B126} (1977) 298; Yu.L. Dokshitser:{\it
Sov. Phys. JETP} {\bf 46} (1977) 641.
\bibitem{OPE}
K. Wilson: {\it Phys. Rev.} {\bf 179} (1969) 1444.
\bibitem{FACT}
 J.C. Collins, D.E. Soper and G. Sterman:{\it Nucl.
Phys.}{\bf B308} (1988) 833.
\bibitem{FORTE}
S.Forte: plenary talk theory DIS97 {\it these proceedings}.
\bibitem{BFKL}
E.A. Kuraev,  L.N. Lipatov and V.S. Fadin: {\it Sov. Phys. JETP} {\bf 45}
        (1977) 199 ;  \,\,Ya.Ya. Balitskii and L.V. Lipatov: {\it Sov. J.
Nucl. Phys.} {\bf 28} (1978) 822;\,\,L.N. Lipatov: {\it Sov. Phys. JETP}
{\bf 63} (1986) 904.
\bibitem{CCFM}
M. Ciafaloni: {\it Nucl. Phys.} {\bf B296} (1987) 249; S. Catani, F.
Fiorani and G. Marchesini: {\it Phys. Lett.} {\bf B234} (1990) 339,{\it
Nucl. Phys.} {\bf B336} (1990) 18.
 \bibitem{MU94}
A.H. Mueller: {\it Nucl. Phys.}{\bf B415}(1994)373.
\bibitem{FADIN}
V.Fadin: {\it these proceedings}.
\bibitem{CIAFALONI}
G. Camisi and M. Ciafaloni:   hep - ph /9701303 and references therein.
\bibitem{BRA}
M. Braun:{\it Phys. Lett.} {\bf B345} (1995) 155 ;E.Levin:{\it Nucl.
Phys.} {\bf B 453} (1995) 303.
\bibitem{SALAM}
G.Salam:{\it these proceedings}.
\bibitem{WHITE}
A.White:{\it these proceedings}.
\bibitem{DELDUCA}
V. Del Duca: minireview,{\it these proceedings}.
\bibitem{BALL}
R.D. Ball and S. Forte:hep-ph/9703417; R.D. Ball:{\it these proceedings}.
\bibitem{BALITSKY}
I. Balitsky: ; {\it these proceedings}.
\bibitem{KOVNER}
 J. Jalilian-Marian, A. Kovner, A. Leonidov and H. Weigert: hep -
ph/9701284; A. Kovner:{\it these proceedings}.
\bibitem{BAWU}
J. Bartels and M. Wusthoff: {\it Z. Phys.} {\bf C66} (1995) 157.
\bibitem{MU97}
A.H. Mueller: {\it Phys. Lett.} {\bf B396} (1997) 251.
\bibitem{LEVIN}
E. Gotsman, E. Levin and U. Maor: hep -ph/9705205, {\it Phys. Lett. (in
press)}; E. Levin: {\it these proceedings}.
 \bibitem{RY}
M.G. Ryskin: {\it Z. Phys.}{\bf C57}(1993) 89.
\bibitem{FIVE}
S.J. Brodsky,L.L. Frankfurt, J.F. Gunion, A.H. Mueller  and M. Strikman :
{\it Phys. Rev.}{\bf D50}(1994)3134.
\bibitem{CFS}
J.Collins,L.Frankfurt and M. Strikman: hep - ph /9611433.
\bibitem{BL}
S.J. Brodsky and G.P.Lepage: {\it Phys. Rev.} {\bf D22} (1980) 2157.
\bibitem{BALO}
J. Bartels and M. Loewe: {\it Z. Phys.} {\bf C12} (1982) 263.
\bibitem{GLR} 
L.V. Gribov, E.M. Levin and M.G. Ryskin: {\it Phys. Rep.} {\bf 100} (1983)
1.
\bibitem{GER}
F.-M. Dittes,D.Muller,D.Raschik,B.Geyer and J.Horeijsi: {\it Fortschr.
Phys.} {\bf 42} (1994) 101.
\bibitem{RAD}
A.V. Radyushkin: {\it Phys. Lett.}{\bf B380} (1996) 417, {\bf B385} (1996)
333,hep - ph/9704207,{\it these proceedings}.
\bibitem{FRE}
L. Frankfurt, A. Freund, V. Guzey and M. Strikman: hep - ph
/9703449; A.Freund:{\it these proceedings}.
\bibitem{ELLISHT}
R.K.Ellis,W. Furmanski and R. Petronzio: {\it Nucl.Phys.} {\bf B207}
(1982)1, {\bf B212} (1983) 29.
\bibitem{BULI}
A.P. Bukhvostov,G.V. Frolov, L.N. Lipatov and E.A. Kuraev: {\it Nucl.
Phys.} {\bf B258} (1985) 601.
\bibitem{HTLOWX}
J. Bartels: {\it Phys. Lett.} {\bf B298} (1993) 204, {\it Z. Phys.} {\bf
C60} (1993) 471; E.M. Levin, M.G. Ryskin and A.G. Shuvaev: {\it Nucl.
Phys.} {\bf B387} (1992) 589.
\bibitem{BARTELSHT}
J. Bartels: {\it these proceedings}.
\bibitem{MUQI}
A.H. Mueller and J. Qiu: {\it Nucl. Phys.}  {\bf B268} (1986) 427.
\bibitem{AGL}  
A.L. Ayala, M.B. Gay Ducati and E.M. Levin:  hep-ph
9604383, {\it Nucl. Phys.}{\bf B493}(1997) 305.
\bibitem{WEIGERT}
 J. Jalilian-Marian, A. Kovner, L. McLerran  and H. Weigert:
hep - ph /9606337; H. Weigert: {\it these proceedings}.
\bibitem{MU90}
A.H.Mueller: {\it Nucl. Phys.} {\bf B335} (1990) 115.
\bibitem{KTFACT}
S.Catani,M. Ciafaloni and F. Hautmann: {\it Phys. Lett.} {\bf B242}
(1990) 97, {\it Nucl. Phys.} {\bf B366} (1991) 135; J.C. Cillins and R.K.
Ellis: {\it Nucl. Phys.} {\bf B360} (1991) 3; E.M. Levin,M.G.
Ryskin,Yu.M.Shabelsky and A.G. Shuvaev: {\it Sov.J.Nucl.Phys.} {\bf 53}
(1991) 657.
\bibitem{BJ}
J.D.Bjorken,J. Kogut and D. Soper: {\it Phys. Rev.} {\bf D3} (1971) 1382;
J.D.Bjorken and J. Kogut: {\it Phys. Rev.} {\bf D8} (1973)  341.
\bibitem{NACHT}
O. Nachtmann: {\it Ann. Phys.} {\bf 209} (1991) 936.
\bibitem{BUCH}
W. Buchmuller, M.F. McDermott and A. Hebecker:  hep - ph/9703314 and
reference therein;W.
Buchmuller: {\it these proceedings}.
\bibitem{LMRT}
E.M. Levin, A.D. Martin, M.G. Ryskin and T. Teubner: hep - ph /966606443,
{\it Z. Phys. ( in press)}.
\bibitem{MCLV}
L. McLerran and R. Venugopalan: {\it Phys. Rev.} {\bf D49} (1994)
2233,3352, {\bf D50} (1994) 2225, {\bf D53} (1996) 458.
\bibitem{LALE}
E. Laenen and E. Levin: {\it Ann. Rev. Nucl. Part.} {\bf 44} (1994) 199.
\bibitem{KOVCH}
Yu.V. Kovchegov: {\it Phys. Rev.} {\bf D54} (1996) 5463, hep -
ph/9701229; Yu.V. Kovchegov, A.H. Mueller and S. Wallon: hep -
ph/9704369.
\bibitem{BENEKE}
M. Beneke: minireview, {\it these proceedings}.
\bibitem{AKHOURY}
M.Akhoury: {\it these proceedings}.
\bibitem{MARCHESINI}
G. Marchesini: {\it these proceedings}.
\bibitem{STERMAN}
G. Sterman: {\it these proceedings}.
\bibitem{SOT}
A. Sotiropoulos: {these proceedings}.
\bibitem{SCHREMPT}
F. Schrempp: minireview, {\it these proceedings}.
\bibitem{LATTICE}
G. Schierholz: minireview, {\it these proceedings}.
\bibitem{THORNE}
R.S.Thorne: hep - ph /9701241; J. Kwiecinski, A.D. Martin and A.M.
Stasto: hep - ph /9703445.
\end{thebibliography}
 \end{document}